# Intrinsic Spin Seebeck Effect in Au/YIG

D. Qu[1], S. Y. Huang[1,2], Jun Hu[3], Ruqian Wu[3], and C. L. Chien[1]*

**Affiliations:**

[1]Department of Physics and Astronomy, Johns Hopkins University, Baltimore Maryland 21218, USA

[2]Department of Physics, National Tsing Hua University, Hsinchu 300, Taiwan

[3] Department of Physics and Astronomy, University of California, Irvine, California 92697, USA

*clc@pha.jhu.edu

**Abstract:**

The acute magnetic proximity effects in Pt/YIG compromise the suitability of Pt as a spin current detector. We show that Au/YIG, with no anomalous Hall effect and a negligible magnetoresistance, allows the measurements of the intrinsic spin Seebeck effect with a magnitude much smaller than that in Pt/YIG. The experiment results are consistent with the spin-polarized density-functional calculations for Pt with a sizable and Au with a negligible magnetic moment near the interface with YIG.

**PACS numbers**: 72.15.Jf, 72.20.Pa, 85.80.-b, 85.75.-d

The exploration of spintronic phenomena has been advanced towards the manipulation of a pure spin current without a charge current. A pure spin current can be realized by compelling electrons of opposite spins to move in opposite directions, or be carried by spin waves (magnons). Pure spin current is beneficial for spintronic operations with the attributes of maximal angular momentum and minimal charge current thus with much reduced Joule heating, circuit capacitance and electromigration. In the spin Hall effect (SHE), a charge current driven by a voltage gradient can generate a transverse spin current [1]. Using the spin Seebeck effect (SSE), a temperature gradient can also generate a spin current. Consequently, the SSE, within the emerging field of "spin caloritronics", where one exploits the interplay of spin, charge, and heat, has attracted much attention. SSE has been reported in a variety of ferromagnetic (FM) materials (metal [2], semiconductor [3], or insulator [4]), where the pure spin current is detected in the Pt strip patterned onto the FM material by the inverse spin Hall effect (ISHE) with an electric field of $\boldsymbol{E}_{SHE} = D_{ISHE}\boldsymbol{j}_S \times \boldsymbol{\sigma}$, where $D_{ISHE}$ is the ISHE efficiency, $\boldsymbol{j}_S$ is the pure spin current density diffusing into the Pt strip and $\boldsymbol{\sigma}$ is the spin direction. Consider a FM layer sample in the $xy$-plane, there are two ways to observe SSE using either the transverse or the longitudinal SSE configuration with a temperature gradient applied either in the sample plane ($\nabla_x T$) or out-of-plane ($\nabla_z T$) respectively. Various potential applications of SSE have already been proposed [5].

However, the SSE is not without controversies and complications. One fundamental mystery is that SSE has been reported in macroscopic structures on the mm scale whereas the spin diffusion length within which the spin coherence is preserved is only on the nm scale [2-4]. Furthermore, the previous reports of SSE have other unforeseen complications. In the transverse geometry of SSE with an intended in-plane $\nabla_x T$, due to the overwhelming heat conduction through the substrate, there exists also an out-of-plane $\nabla_z T$, which gives rise to the anomalous Nernst effect (ANE) with an electric field of $\boldsymbol{E_{ANE}} \propto \nabla_z T \times \boldsymbol{m}$, where $\boldsymbol{m}$ is the magnetization direction [6]. The ANE is very sensitive in detecting $\nabla_z T$ in a manner similar to the high sensitivity of the anomalous Hall effect to perpendicular magnetization in FM layers less than 1 nm in thickness. As a result, $\boldsymbol{E_{SHE}} \propto \boldsymbol{j_S} \times \boldsymbol{\sigma}$ due to SSE with $\boldsymbol{j_S}$ in the z-direction and $\boldsymbol{E_{ANE}} \propto \nabla_z T \times \boldsymbol{m}$ due to ANE are both along the y-direction and asymmetric in magnetic field. The voltages of SSE due to $\nabla_x T$ and ANE due to $\nabla_z T$ are additive, entangled, and inseparable [6].

In the longitudinal SSE using Pt on a FM insulator (e.g., YIG), while the temperature gradient $\nabla_z T$ is unequivocally out-of-plane, one encounters a different issue of magnetic proximity effects (MPE) in Pt in contact with a FM material. As a result, in the longitudinal configuration there is also entanglement of SSE and ANE [7]. These complications, when present, prevent the unequivocal establishment of SSE in either the transverse or the longitudinal configuration. The characteristics of the intrinsic SSE including its magnitude, remain outstanding and unresolved issues.

In this work, we report the measurements of intrinsic SSE in gold (Au) using the longitudinal configuration with an unambiguous out-of-plane ($\nabla_z T$) gradient near room temperature. It is crucial to identify metals other than Pt that can unequivocally detect the pure spin current without MPE. Gold offers good prospects since it has been successfully used as a substrate or underlayer for ultrathin magnetic films. We use polished polycrystalline yttrium iron garnet (YIG=$Y_3Fe_5O_{12}$) a well-known FM insulator with low loss magnons as the substrate. A large spin-mixing conductance at Au/YIG interface has been reported using ferromagnetic resonance [8]. Our results show that Au(*t*)/YIG does not have the large anomalous Hall effect and large MR that plagued Pt(*t*)/YIG but exhibits an unusual thickness dependence in the thermal transport. These results allow us to place an upper limit for the intrinsic SSE of about 0.1 μV/K much smaller than the thermal effect in Pt/YIG [7].

Thin Au films have been made by magnetron sputtering on YIG and patterned into parallel wires and Hall bars. As shown in the inset of Fig. 1(a), the *xyz* axes are parallel to the edges of the YIG substrates. The parallel wires with ascending order of thickness from 4 nm to 12 nm are in the *xy*-plane and oriented in the *y*-direction, where each wire is 4 mm long, 0.1 mm wide, and 2 mm apart. The Hall bar samples (inset of Fig. 1b) consist of one long segment along the *x*-direction and several short segments along the *y*-direction. For the MR measurements current is along the *x*-axis, for the thermal transport measurement $\nabla_z T$ is along the *z*-axis, and the magnetic field is in the *xy*-plane in both cases. We use 4-probe and 2-probe measurements for MR and thermal voltage respectively. The multiple wires facilitate a systematic study of the thickness dependence of electric transport and thermal measurement under the same uniform thermal gradient with a temperature difference of $\Delta T_z \approx$ 10 K. The sample was sandwiched between, and in thermal contact with, two large Cu blocks kept at constant temperatures differing by 10 K.

We first describe the thickness dependence of electrical resistivity (*ρ*) of the Au wires. As expected *ρ* increases with decreasing film thickness as shown in Fig. 1(a). The results can be well described by a semiclassical theoretical model in the frame of Fuchs-Sondheimer (FS) theory [9], which includes the contributions from thickness ($t$) as well as surface scattering ($p$) and grain boundary scattering ($\xi$), $\rho = \rho_\infty\{1-(1/2+3\lambda/4t)[1-p\exp(-t\xi/\lambda)]\exp(-t/\lambda)\}^{-1}$ for $t/\lambda > 0.1$. Using bulk resistivity ($\rho_\infty = 2.2$ μΩcm) and the mean free path ($\lambda=37$ nm) [10], we find the data can be well described by $p = 0.89$ and $\xi = 0.37$ as shown by the solid line in Fig. 1(a).

The anomalous Hall effect (AHE) is an essential measurement for assessing MPE. Hall measurements of the Au/YIG Hall bar samples have been made from 2K to 300K as shown in Fig. 1(b). The Hall resistance of Au/YIG is linear in magnetic field at all temperatures (2-300 K) showing only the ordinary Hall effect (OHE) with no observable AHE. In contrast, strong AHE has been observed in Pt/YIG due to the acute MPE [7]. The Hall constant ($R_H = 1/ne$) of Au/YIG indicates the carrier concentration $n \approx 6\times10^{22}$ cm$^{-3}$ as shown in Fig. 1(c), essentially constant from 2 K to 300 K, is in good agreement with the bulk carrier concentration of $n = 5.9\times10^{22}$ cm$^{-3}$ [11]. The spin polarized moment induced in Au, is very small, if any, i.e., Au is not appreciably affected by MPE and will be further discussed below.

We employ the longitudinal configuration with spin current along the out-of-plane temperature gradient $\nabla_z T$ to determine the thickness dependence of the thermal transport of Au/YIG, and to compare the results with those of the Pt/YIG. As shown in Fig. 2(a), the transverse thermal voltage (in the $y$-direction) across the Au strip is asymmetrical when the magnetic field is along the $x$-axis with the same sense as that for the Pt strip. The same sign of the thermal spin-Hall voltage between Pt and Au is consistent with the theoretical calculation of positive values of spin Hall conductivity in Pt and Au [12].

However, there are several distinct differences between the thermal results of Au/YIG and Pt/YIG. We take $\Delta V_{th}$ as the magnitude of spin-dependent thermal voltage between the positive and the negative switching fields. As shown in Fig. 2(b), the value of $\Delta V_{th}$ of the Pt(*t*)/YIG is far larger, increasing sharply and unabatedly with decreasing *t* to a value of 64 μV at *t* = 2.2 nm, due to the strong MPE at the interface between Pt and YIG. In contrast, the thermal voltage $\Delta V_{th}$ of the Au/YIG samples is much smaller than that of Pt/YIG and it varies with thickness (*t*) in a non-monotonic manner as shown in Fig. 2(c). The value of $\Delta V_{th}$ is negligible (less than 0.2 μV) for $t \leq 7$ nm, increasing to a maximum of 1.3 μV at *t* = 8 nm before decreasing at larger thicknesses. This contrasting behavior shows that there is much smaller, perhaps negligible, MPE in Au/YIG. Consequently, the measured thermal voltage may be attributed entirely to intrinsic SSE. With a maximal $(\Delta V_{th})_{max} \approx 1.3$ μV at *t* = 8 nm at ΔT of 10 K, the strength of the intrinsic SSE in Au/YIG is about 0.1 μV/K, far smaller than the values in Pt(*t*)/YIG of 6 μV/K at *t* = 2.2 nm, and 1 μV/K at *t* = 10 nm, by one to two orders of magnitude. This suggests most of the spin-dependent thermal voltage in Pt/YIG is due to ANE and *not* SSE. From the value of $S_{xy} \approx 6 \times 10^{-3}$ μV/K ($S_{xy}=E_{xy}/\nabla T=(\Delta V_{th}/l)/(\Delta T/d)$, where $\Delta V_{th}$ is the thermal voltage, *l* is the distance between the voltage leads, $\Delta T$ is the temperature difference and *d* is the thickness of Au/YIG sample) we measured and using the Seebeck coefficient $S_{xx} \approx 1.9$ μV/K of Au at 300 K [13], we obtain a spin Nernst angle of $\Theta_N = S_{xy}/S_{xx} \approx 0.003$, which is very close to the spin Hall angle $\Theta_H = 0.0016$, defined as the ratio of spin Hall and charge conductivities, from spin pumping measurement [14].

However, we have observed MR, albeit with very small but clear signals, in Au($t$)/YIG. The MR result of Au(7 nm)/YIG Hall bar sample is shown in Fig. 3(a). It is of a very small magnitude of $\Delta\rho/\rho \approx -4 \times 10^{-6}$, where $\Delta\rho = \rho_{\|} - \rho_T$, about two orders of magnitude smaller than those of Pt($t$)/YIG as shown in Fig. 3(b). Nevertheless all the Au($t$)/YIG with 4 nm $\leq t \leq$ 11 nm show similarly small but measurable $\Delta\rho/\rho$. More unexpectedly, the MR of Au($t$)/YIG has the *opposite* angular dependence as that of the usual anisotropic MR (AMR). In the AMR of most *3d* ferromagnetic metals of Fe, Co, Ni, and their alloys, the common behavior is positive $\Delta\rho$, that is $\rho_{\|} > \rho_T$, the resistivity with current parallel to, is higher than that with the current perpendicular, to the magnetization aligned by a magnetic field. The MR observed in Pt($t$)/YIG also has the same behavior of $\Delta\rho > 0$. In contrast, the small MR in Au($t$)/YIG is opposite with $\rho_T > \rho_{\|}$, or inverse AMR, The mechanism of this up behavior in Au($t$)/YIG is not yet fully understood, but probably due to spin-dependent scattering at interface between Au and YIG, supported by the fact that $|\Delta\rho|$ increases with decreasing Au films thickness. One notes that inverse AMR has occasionally been reported in thin Co films. The s-d scattering influenced by spin-orbital and electron-electron interactions may be enhanced by the disorder in thin films [15].

To assess the magnetic moments of Pt and Au near the interface with YIG, spin-polarized density functional calculations have been carried out with the Vienna ab initio simulation package (VASP), [16,17] at the level of the generalized gradient approximation (GGA) [18] with a Hubbard U correction for Fe-3d orbitals in YIG. We use the projector augmented wave (PAW) method for the description of the core-valence interaction [19,20]. The YIG structure has two Fe sites: tetrahedral $Fe_t$ and octahedral $Fe_o$.

To model the Pt/YIG and Au/YIG interfaces, we construct a superlattice structure with a slab of YIG(111) of about 6 Å thick along with a 4-layer Pt or Au film of about 7 Å thick. In the initial configuration, the $Fe_o$ atoms match the hcp sites of Pt(111) or Au(111) slab. During the relaxation process, the in-plane lattice constant has been fixed at the experimental value of the bulk YIG, with a dimension of $17.5 \times 17.5$ Å$^2$, and thickness of superlattice [notated as c in Fig. 4(a)] is allowed to change. All atoms are fully relaxed until the calculated force on each atom is smaller than 0.02 eV/Å. For this large unit cell with 274 atoms, we find that a single Γ point is enough to sample the Brillouin zone. The optimized atomic structure of Pt/YIG in Fig. 4(a) shows significant reconstructions in both Pt and YIG layers. The average bond lengths are: $d_{Pt-O}$ ~ 2.2 Å, $d_{Pt-Fe}$ ~ 2.6 Å. Au/YIG has a similar structure.

It is important to note that all four Pt layers are significantly spin polarized as shown in Fig. 4(b). The Pt layers adjacent to the interfaces [labeled by 1 and 4 in Fig. 4(b)] tend to couple ferromagnetically to their neighboring Fe atoms in YIG, as found in most studies for Pt on magnetic substrates. The local spin moments of Pt atoms in the $Pt_2$ and $Pt_3$ layers can still be as large as 0.1 $\mu_B$. By integrating spin density in the lateral planes, we can obtain the z-dependent spin density as shown in Fig. 4(c). Clearly, the spin polarization in all Pt layers is significant for the measurement of SSE. In particular, the total spin moments of the $Pt_2$ and $Pt_3$ layers (each has 36 Pt atoms) are about 0.8 $\mu_B$ and 1.1 $\mu_B$, respectively, even after the mutual cancelation with the intra-layer antiferromagnetic ordering. In contrast, spin polarizations induced in the Au layers are much weaker, with the maximum local spin moment smaller than 0.05 $\mu_B$ and the integrated spin moment in the entire Au layers smaller than 0.1 $\mu_B$. Therefore, one can view Au as nearly "nonmagnetic" in contact with YIG, in contrast to Pt.

The sizable magnetic moments of Pt near the interface from the theoretical calculations is consisting with the strong MPE shown in Pt($t$)/YIG by the electric transport. Therefore, the ANE and SSE are not only entangled but with ANE dominating in Pt/YIG. In contrast, the negligible Au moments from theoretical calculations is also consistent with no apparent AHE in Au($t$)/YIG. The only noticeable magnetic characteristic is the inverse AMR of Au($t$)/YIG but with a magnitude two orders smaller than that of Pt/YIG. This indicates that most, if not all, of the thermal voltage measured in Au/YIG is due to the intrinsic SSE as a result of the pure spin current injected from YIG.

As shown in Fig. 2(c), the measurement of the thickness dependence is essential in revealing the non-monotonic dependence of intrinsic SSE voltage in Au/YIG due to the spin diffusion length $\lambda_{SF}$. For very thin Au layer with t < 6 nm, $\lambda_{SF}$ is short due to the large resistivity from interface and boundary scattering, thus no appreciable spin current could survive intact, and this results in negligible $\Delta V_{th}$. As the Au film thickness increases, the value of $\Delta V_{th}$ exhibits a rapid rise reaching a maximum of 1.3μV at $t$~8 nm and then decreases owing to the spin flip relaxation mechanism. Using the expression $l_{sf}=\pi/(2k_F^2)\sqrt{(3/2)}\ h/e^2\sqrt{(\tau_{sf}/\tau)}\ \sigma_{xx}$ including the Fermi wave vector $k_F$, the conductivity $\sigma_{xx}$, the mean time between collisions $\tau$ and the mean time between spin-flipping collision $\tau_{sf}$, we estimate $\lambda_{SF} \approx 40$ nm. [21] The critical thickness of 8 nm is close to spin diffusion length 10.5 nm evaluated from weak localization [10]. Given the weak inverse AMR and the nonexistent AHE, the thermal signal of 0.1 μV/K measured in Au/YIG at an optimal thickness of 8 nm should be considered as an upper limit of the intrinsic SSE effect. The spin Hall angle between Au and YIG might be further enhanced by chemical modification on the YIG surface at high temperature. But a careful surface treatment is very important to avoid the metallic state of Fe developed, which could result in a reduction of spin mixing conductance and contamination in SSE [22].

In summary, we use Au(t)/YIG with no anomalous Hall signals and a very weak inverse MR results with non-monotonic dependence of spin-thermal voltage to show that the acute magnetic proximity effects that plagued Pt/YIG do not affect Au/YIG. The thermal voltage in Au/YIG is thus due to primarily intrinsic spin Seebeck effect with an upper limit of 0.1 µV/K. Although the spin Hall angle of Au is smaller than that of Pt, Au is a good spin current detector, far better than Pt.

**Acknowledgments:** The work is supported at Johns Hopkins University by US NSF (DMR 05-20491) and Taiwan NSC (99-2911-I-007- 510), and at University of California by DOE-BES (Grant No: DE-FG02-05ER46237) and by NERSC for computing time.

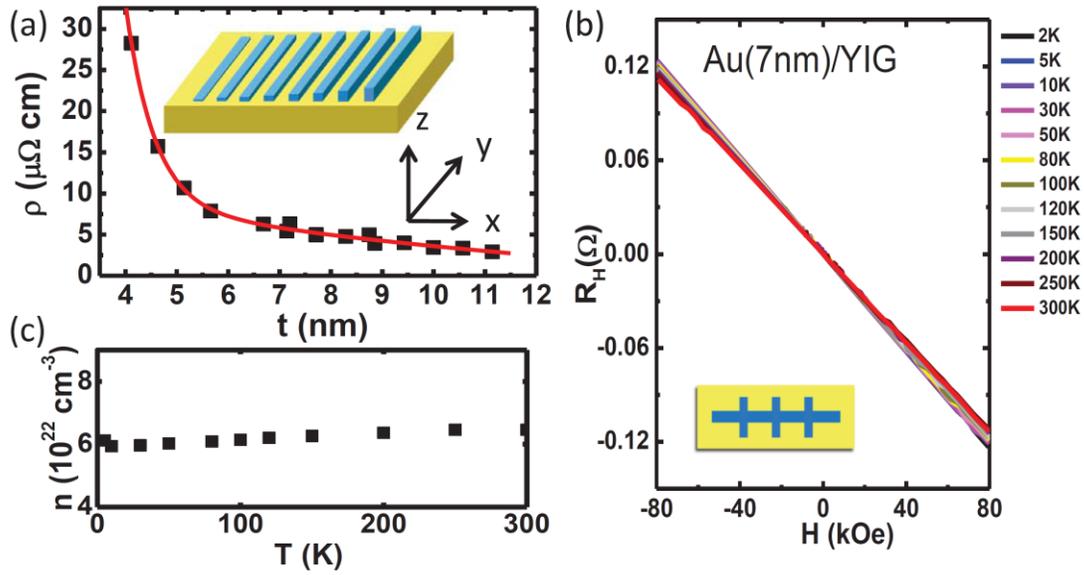

**Fig. 1** (color online). (a) Resistivity as a function of Au thickness *t* for Au /YIG. The solid line represents semiclassical theoretic fittings. Inset is schematic diagram of multiple patterned strips with ascending thickness. (b) Field dependence of Hall resistance $R_H$ at different temperatures for Au(7 nm)/YIG. Inset is schematic diagram of patterned Hall bar. (c) Carrier concentration as a function of temperature for Au(7 nm)/YIG.

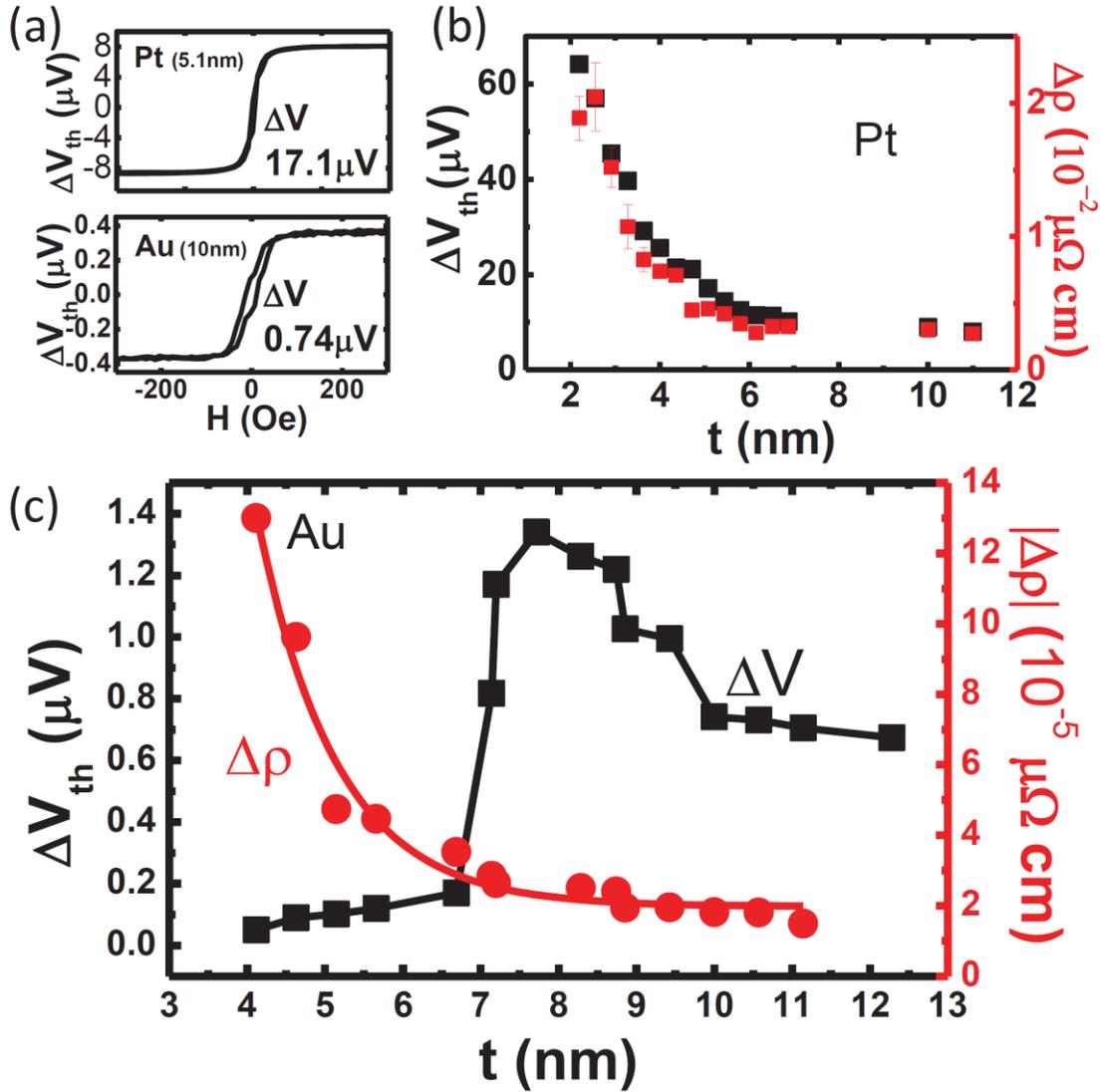

**Fig. 2** (color online). (a) Field-dependent thermal voltage for Pt(5.1nm)/YIG and Au(10nm)/YIG. Thermal voltage (left scale) and $\Delta\rho$ (right scale) for multiple strips as a function of Pt thicknesses (b) and Au thicknesses (c) on YIG. All thermal results are under a temperature difference of $\Delta T \approx 10$ K.

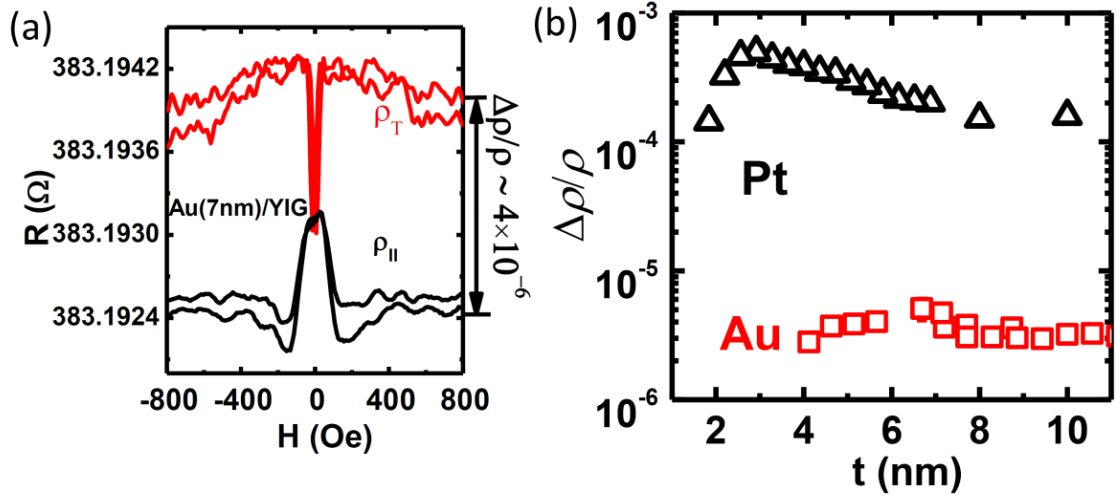

**Fig. 3** (color online). (a) Magnetoresistance (MR) result of Au(7nm)/YIG as a function of magnetic field H at $\theta = 0°(\rho_{\parallel})$ and $90°$ $(\rho_T)$. (b) AMR ratio as a function of metal layer thickness $t$ for Pt/YIG (open triangles) and Au/Pt (open squares).

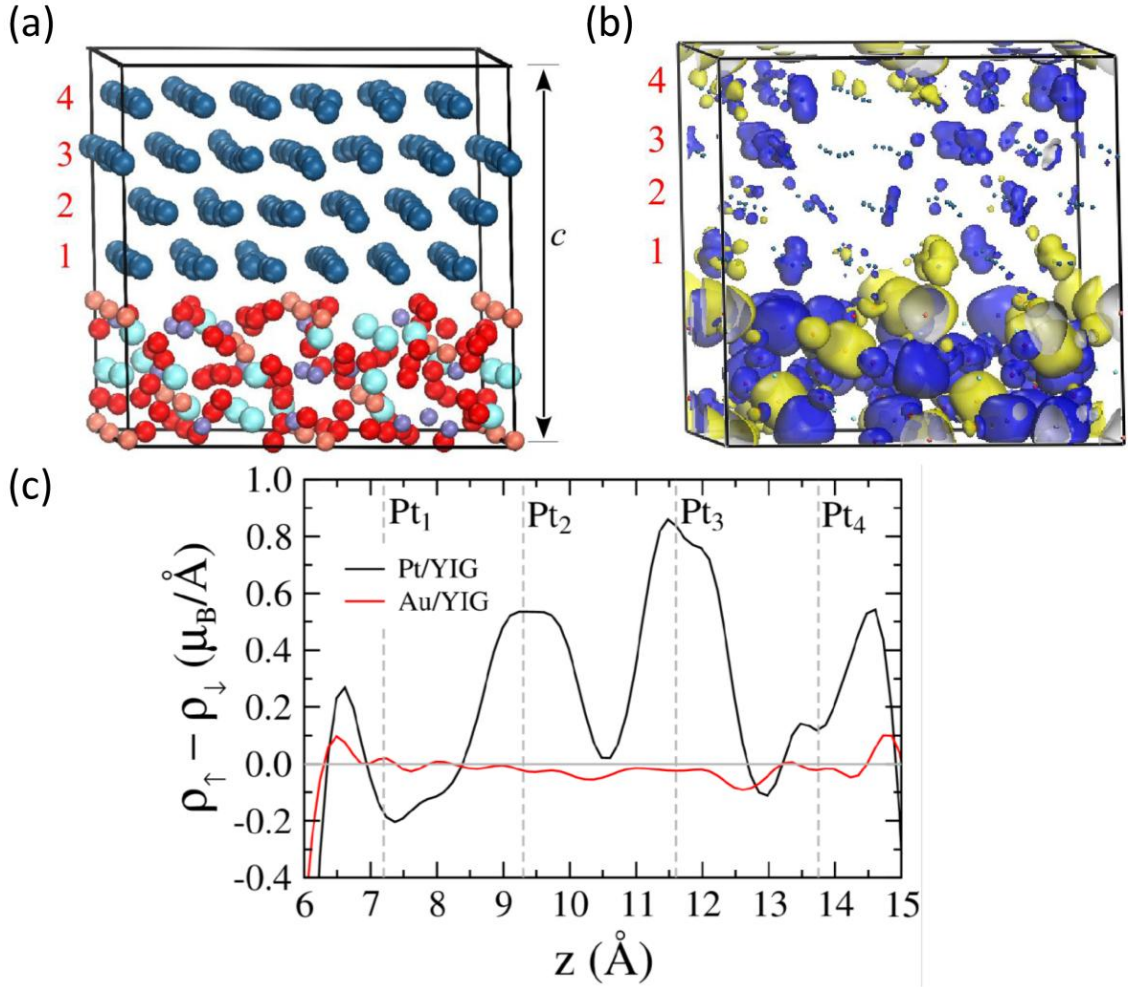

**Fig. 4** (color online). (a) The optimized structural model of Pt/YIG. The teal, coral, purple, cyan and red spheres represent for Pt, $Fe_o$ (center of octahedron), $Fe_t$ (center of tetrahedron), Y and O atoms, respectively. The thickness of the superlattice, denoted as $c$, is 15.6 Å after relaxation. The numerals in the left side label the Pt layers for the convenience of discussions. (b) Isosurfaces of spin density (at 0.03 e/ Å$^3$) of Pt/YIG. The blue and yellow isosurfaces are positive and negative spin polarizations. (c) Planar averaged spin density along $c$ axis. The vertical dashed lines indicate the average z-coordinates of Pt and Au layers. Arrows ↑ and ↓ stand for the majority spin and minority spin contributions, respectively.